\documentclass[namedreferences]{solarphysics}
\usepackage[hyperref,optionalrh,solaromanenum]{spr-sola-addons} 
\usepackage{graphicx}                    
\usepackage{amssymb}                    

\begin{document}
\begin{article}
\begin{opening}

\title{A Dynamo-Based Prediction of Solar Cycle 25}
%

\author[addressref={aff1,aff2}]{\fnm{Wei}~\lnm{Guo}}
\author[addressref={aff1,aff2},corref,email={jiejiang@buaa.edu.cn}]{\inits{J.}\fnm{Jie}~\lnm{Jiang}}\orcid{0000-0001-5002-0577}
\author[addressref={aff3,aff4}]{\fnm{Jing-Xiu}~\lnm{Wang}}
\address[id=aff1]{School of Space and Environment, Beihang University, Beijing, China}
\address[id=aff2]{Key Laboratory of Space Environment Monitoring and Information Processing of MIIT, Beijing, China}
\address[id=aff3]{School of Astronomy and Space Science, University of Chinese Academy of Sciences, Beijing, China}
\address[id=aff4]{Key Laboratory of Solar Activity, National Astronomical Observatories, Chinese Academy of Sciences, Beijing 100101, China}

%
\runningauthor{Guo et al.}
\runningtitle{Solar Cycle 25 Prediction}


\begin{abstract}
  Solar activity cycle varies in amplitude. The last Cycle 24 is the weakest in the past century. Sun's activity dominates Earth's space environment. The frequency and intensity of the Sun's activity are accordant with the solar cycle. Hence there are practical needs to know the amplitude of the upcoming Cycle 25. The dynamo-based solar cycle predictions not only provide predictions, but also offer an effective way to evaluate our understanding of the solar cycle. In this article we apply the method of the first successful dynamo-based prediction developed for Cycle 24 to the prediction of Cycle 25, so that we can verify whether the previous success is repeatable. The prediction shows that Cycle 25 would be about 10\% stronger than Cycle 24 with an amplitude of 126 (international sunspot number version 2.0). The result suggests that Cycle 25 will not enter the Maunder-like grand solar minimum as suggested by some publications. Solar behavior in about four to five years will give a verdict whether the prediction method captures the key mechanism for solar cycle variability, which is assumed as the polar field around the cycle minimum in the model.
\end{abstract}

%

\end{opening}

%
\section{Introduction}\label{s:Introduction}
\sloppy{}
Solar Cycle 24 has been the weakest cycle during the past century. Before its start, there were 265 and 262 spotless days in the years 2008 and 2009, respectively \footnote{http://www.sidc.be/silso/spotless}. Around the end of Cycle 24, the spotless days increased strikingly again. There were 208 and 274 spotless days in the years 2018 and 2019, respectively. Long-term spotless days intrigue the high interest of general public and scientific communities regarding the amplitude of Cycle 25. Some studies even claim that there will be a Maunder-like grand minimum in the near future \citep[e.g.,][]{Morner2015,Zharkova2020}. Furthermore, Sun's activity dominates Earth's space environment, which is essential for our technological society. The frequency and intensity of Sun's activity are accordant with the solar cycle. Hence there are practical needs to know the future solar cycle amplitude as early as possible.

People have started to predict solar cycles since they were aware of the periodic behaviour of solar activity. Existing attempts to predict future solar cycle(s) can be broadly divided into three types \citep{Petrovay2020}: (1) extrapolation methods, deriving a prediction from a purely mathematical analysis of the solar activity time series, (2) precursor methods based on correlations between certain measured quantities in the declining phase of a cycle and the strength of the next one, and (3) physics-based methods. The method we present in this article belongs to the last type. For the prediction of Cycle 24, \cite{Pesnell2008} well illustrated the wide divergence of predictions covering almost the entire range of possible values of the peak sunspot number (from 50 to 190 based on the sunspot number data Version 1.0). Recently \cite{Nandy2021} summarized predictions for Solar Cycles 24 and 25.

Understanding solar cycle mechanisms and further applying the results to predict an upcoming cycle remain an important aspect of research in solar-terrestrial physics and astrophysics. Physics-based predictions belong to such kind of research and also provide an effective way to verify our understanding in return. The ability to provide a testbed for dynamo models makes physics-based prediction methods distinct from the other two types. A physics-based prediction consists of two key ingredients. One is the quantitative model, which is assumed to capture the key ingredients describing the solar cycle. The other is the appropriately observed data assimilated into the model.

Babcock-Leighton (BL) type flux transport dynamo (FTD) is the workhorse for understanding the solar cycle \citep{Karak2014}. A key ingredient of the models is the BL mechanism, which was first proposed by \cite{Babcock1961} and further elaborated by \cite{Leighton1969}. The essence of the BL mechanism is that the poloidal field is generated by emergence and decay of tilted sunspot groups over the solar surface. During the Cycle 23 minimum, the first generation of physics-based prediction methods came out by two groups with the progress of the type of dynamo in understanding the solar cycle. However, the two groups gave strikingly different predictions. Group I, i.e., \cite{Dikpati2006a} and \cite{Dikpati2006b}, predicted that Cycle 24 would be 30\%-50\% stronger than Cycle 23. Group II, i.e., \cite{Choudhuri2007} and \cite{Jiang2007}, predicted that Cycle 24 would be about 30\% weaker than Cycle 23. For a long time, people have only attributed the divergence to the flux transport parameters of the dynamo models, which correspond to the so-called advection-dominated and diffusion-dominated dynamos \citep{Yeates2008}. Section 4 of \cite{Karak2014} gave details of each models' differences. The progresses in understanding the solar cycle variability during the past decade have further clarified another cause of the divergence, which is the observational data assimilated into each model. The data assimilated by Group I and II are the sunspot group areas and the surface poloidal field, respectively.

Since the emergence of the first generation of physics-based prediction, major progresses in the understanding of the solar cycle in the framework of the BL-type dynamo were made in three aspects. First, the BL mechanism received more evidences and recognition that it actually is at the essence of the solar cycle. \cite{Jiang2013} investigated the behavior of a BL dynamo with a poloidal source term that was based on the observed sunspot areas and tilts for cycles 15-21. They found that the toroidal flux at the base of the convection zone is highly correlated with the maxima of solar activity levels. \cite{Cameron2015} demonstrated that the net toroidal magnetic flux generated by differential rotation within a hemisphere of the convection zone was determined by the emerged magnetic flux at the solar surface. There are two notable properties of the BL-type dynamo. One is that the surface large-scale field due to sunspot emergence, which is directly observable, works as the seed for the subsequent cycle. The other is that the toroidal and poloidal fields are spatially separated. Some magnetic flux transport mechanisms, e.g. meridional flow \citep{Choudhuri1995}, turbulent diffusion \citep{Chatterjee2004}, turbulent pumping \citep{Cameron2013}, are required to connect the two sources. Hence there is a time delay in the segregated sources \citep{Durney2000,Charbonneau2001,Wilmot2006}. These two properties make the BL-type dynamo usable for physics-based predictions.

Second, observational evidences of nonlinear and random mechanisms have been identified in the evolution of the surface large-scale field, corresponding to the surface poloidal field of the BL-type dynamo. \cite{Dasi2010} first noticed the anti-correlation between the cycle strength and the normalized tilt angle. This is regarded as a nonlinearity of the BL mechanism, i.e., tilt quenching. See also \cite{Cameron2010, Jha2020}. \cite{Jiang2020} proposed that the systematic change in latitude has a similar nonlinear feedback on the solar cycle, i.e., latitudinal quenching as the tilt does. This idea has been demonstrated in a 3D dynamo model by \cite{Karak2020}. Except the systematic change in latitude and tilt of sunspot emergence, significant random components, especially in tilt have also been investigated. \cite{Kitchatinov2011,Olemskoy2013,Jiang2014,Cameron2017} indicated that random components in sunspot emergence cause uncertainty in the surface poloidal field generation, and hence dominate the solar cycle variability. The nonlinear and random mechanisms lead to the non-correlation between the cycle strength and the polar field at the end of the cycle. This is clearly demonstrated by observations \citep{Jiang2007,Cameron2010}. The nonlinear and random mechanisms also indicate that when we assimilate data like the poloidal field source into a dynamo model, we cannot linearly convert the sunspot area into the poloidal field. Nevertheless linear conversion was conducted by Group I. While group II assimilated the observed magnetic butterfly diagram around the cycle minimum, incorporating the nonlinear and random mechanisms suggested above.

Third, the less than one cycle persistence of past cycles memory \citep{Yeates2008} has been widely accepted by the community. Both direct observations of the polar field \citep{Svalgaard2005,Svalgaard2020} and indirect proxy of the polar field, e.g.,the smoothed aa-index \citep{Wang2009} and polar faculae measurements \citep{Munoz2013} show a correlation between the polar field at cycle minimum and the subsequent cycle strength with statistical significance. In the framework of the BL-type dynamo, the process of the poloidal flux transport from the surface to the bulk of convection zone is linear. Shearing of the poloidal magnetic field into a toroidal component is also a quasi-linear process under the effect of differential rotation. The correlation provides a constraint on the poloidal flux transport time, which is shorter than one cycle, from the surface to the toroidal field generation layer. The less than one cycle magnetic memory is also responsible for the predictive ability of the precursor of the solar cycle, i.e., the polar field at the cycle's minimum. In addition, \cite{Hazra2020} demonstrated that inclusion of a mean-field $\alpha$-effect in the framework of a flux transport BL dynamo model leads to additional complexities that may impact memory.

With the progress in understanding the solar cycle, predictions based on BL-type dynamo models were developed in order to predict Cycle 25 \citep[e.g.,][]{Bhowmik2018,Labonville2019}. Surface flux transport (SFT) models including nonlinear and random mechanisms can well reproduce the large-scale field evolution over the surface \citep{Jiang2015, Petrovay2019, Wang2020}. They were also applied to Solar Cycle 25 predictions, especially for predicting the possible polar field strength before the cycle's minimum \citep[e.g.,][]{Cameron2016, Hathaway2016, Iijima2017,Jiang2018}. Figure 3 of \cite{Nandy2021} shows that physics-based predictions for Solar Cycle 25 have converged comparing the results given by the two groups for Cycle 24 predictions though there are still some divergences.

Although observations have indicated that the prediction of Group II for Cycle 24 is correct, people always wonder whether Group II were simply the lucky people whose predictions accidentally turned out to be right and whether they included the correct physics of the solar cycle into their making the prediction. This motivates us to apply the method developed by Group II to Cycle 25 prediction and hence to evaluate the model. Therefore the aim of this article is to make a prediction of Cycle 25 using the dynamo-based prediction model, which was developed to predict Cycle 24 by \cite{Choudhuri2007} and \cite{Jiang2007}. In Section \ref{s:model}, we overview the prediction model and data assimilation. Section \ref{s:results} provides the prediction's result. We conclude in Section \ref{s:Conclusion}.

\section{Dynamo-Based Prediction Model and Data Assimilation}\label{s:model}
\cite{Choudhuri2007} and \cite{Jiang2007} used the same BL-FTD dynamo model. Both assimilated observational data of the Sun's polar magnetic field into the solar dynamo model. \cite{Choudhuri2007} updated the poloidal field of the dynamo model during a particular cycle minimum by a non-latitude-dependent factor, which is equal to the ratio between the observed dipole moment (DM) of different cycles. The DM was measured as the absolute value of the difference between the two polar caps fields. The data assimilation method used by \cite{Choudhuri2007} is a drastic simplification. \cite{Jiang2007} fed values of poloidal field at different latitudes into the theoretical model instead of using only the DM value. Here we follow the method by \cite{Jiang2007} to assimilate observational data into the dynamo model.

\subsection{The Flux Transport Dynamo Model}
The dynamo model we adopt is exactly the same as the standard model in \cite{Jiang2007} for the goal of this article, which is to verify whether the success of the aforementioned prediction model can be repeatable. The standard model in \cite{Jiang2007} originates from \cite{Chatterjee2004}. Here we only give an overview of the two-dimensional FTD model. Equations for the axisymmetric model using spherical coordinates ($r, \theta, \phi$) are
\begin{equation}
 \frac{\partial A}{\partial t}+\frac{1}{s}(\mathbf{v_p}\cdot\nabla)(r\sin\theta A)=\eta_{p}\left(\nabla^{2}-\frac{1}{r^{2}\sin^{2}\theta}\right) A+\alpha B
\end{equation}
\begin{eqnarray}
\frac{\partial B}{\partial t}+\frac{1}{r}\left[\frac{\partial}{\partial
r}(rv_rB)+\frac{\partial}{\partial\theta}(v_\theta B)\right]
=\eta_t\left(\nabla^2-\frac{1}{r^2\sin^{2}\theta}\right)B\nonumber\\
+r\sin\theta(\textbf{B}_p\cdot\nabla)\Omega+\frac{1}{r}\frac{d\eta_t}{dr}\frac{\partial}{\partial
r}(rB),
\end{eqnarray}
where $B(r,\theta,t)\mathbf{e_\phi}$ and $\mathbf{B}_{p}=\nabla\times[A(r,\theta,t)\mathbf{e_\phi}]$ correspond to the toroidal and poloidal magnetic field, respectively. The large-scale flow fields include the meridional flow $\mathbf{v_p}=v_r\mathbf{e_r}+v_\theta\mathbf{e_\theta}$ and differential rotation $\Omega\mathbf{e_\phi}$, both of which are time independent. The turbulent diffusivity in the bulk of the convection zone for the poloidal field $\eta_p$ is much larger than that of the toroidal field $\eta_t$. It is the strong $\eta_p$ that leads to the less than one solar cycle magnetic memory. The $\alpha$-term describes the generation of the poloidal field from the toroidal field \citep{Nandy2002,Chatterjee2004}. It obeys Eq.(\ref{eq:alpha}), which constrains the poloidal field generation near the surface due to the BL mechanism:
\begin{eqnarray}
\alpha(r, \theta)=\alpha_{0} \cos \theta \frac{1}{4}\left[1+\mathrm{erf}\left(\frac{r-r_{1}}{d_{1}}\right)\right] \times\left[1-\mathrm{erf}\left(\frac{r-r_{2}}{d_{2}}\right)\right],
\label{eq:alpha}
\end{eqnarray}
where $r_{1}=0.95 R_{\odot}, r_{2}=R_{\odot}, d_{1}=d_{2}=0.025 R_{\odot}, \alpha_{0}=20 \mathrm{ms}^{-1}$.  When the toroidal field exceeds a critical value $B_c=8\times10^4$ G, we remove a part of $B$ from the bottom of the convection zone to the surface. Then the surface $B$ couples with the surface $\alpha$-effect for the regeneration of the surface poloidal field. The meridional circulation dominates the cycle period of the model, which is 10.84yr. The fixed cycle period of the model causes that the model could only give a reasonable prediction of the cycle amplitude, but not the cycle shape.

Calculations are carried out in a meridional slab $R_b=0.55R_\odot \leq r \leq R_\odot$, $0\leq\theta\leq\pi$. At the poles and the bottom boundary we have $A=0$, $B=0$. At the top, the toroidal field is zero and the poloidal field satisfies a potential field. Observations show that the surface large-scale fields are dominated by the radial component \citep{Svalgaard1978, Sun2011}. Some dynamo models have used the radial boundary conditions to be more consistent with observational constraints \citep[e.g.,][]{Cameron2012,Kitchatinov2018}. Here we keep the boundary conditions the same as the previous model.

The model relaxes to the anti-symmetric periodic solution whatever the configuration of the initial conditions are as demonstrated by \cite{Chatterjee2004}. To keep consistent with the previous prediction model during cycle 23 minimum, the code we use is also Surya \citep{Nandy2002,Chatterjee2004}.

We regard that the poloidal field at solar minimum produced by the final solution is some kind of ``average" poloidal field during a typical solar minimum. The poloidal field during a particular solar minimum may be stronger or weaker than the average field. By assimilating the observed poloidal field during cycle minima into the model, we correct the average poloidal field to make the prediction of the next cycle. A more realistic method is to assimilate the observed poloidal field every Carrington Rotation (CR) as \cite{Jiang2013} did. In order to be consistent with \cite{Jiang2007} and verify its effectiveness in the prediction of the cycle amplitude, we keep the same methodology.

A series of papers by \cite{Hazra2014,Passos2014,Hazra2020} demonstrate that an additional poloidal field source effective on weak fields, e.g., the mean-field $\alpha$-effect driven by helical turbulence \citep{Parker1955}, is necessary for self-consistent recovery of the sunspot cycle from the Maunder-like grand minima. \cite{Xu1982,Eddy1983,Wang2019,Arlt2020} show that there were naked-eye sunspots recorded through the whole Maunder minimum. The naked-eye sunspots indicate that there were sunspots contributing to the polar field evolution even during the Maunder-like grand minima. Our model with only the BL $\alpha$-effect could be still operative for the Maunder-like grand minima.

\subsection{Data Assimilation into the Dynamo Model}
\label{sec:DataAssimilation}
To obtain the observed surface poloidal field at cycle minimum as the input to the dynamo model, we use the Wilcox Solar Observatory (WSO) line-of-sight synoptic charts \footnote{http://wso.stanford.edu/synopticl.html} from May 1976 to the present (from CR1642 to CR2235). A few steps similar to \cite{Jiang2007} have been performed to deal with the data.

\begin{figure}[htbp]
\centering
\includegraphics[scale=0.25]{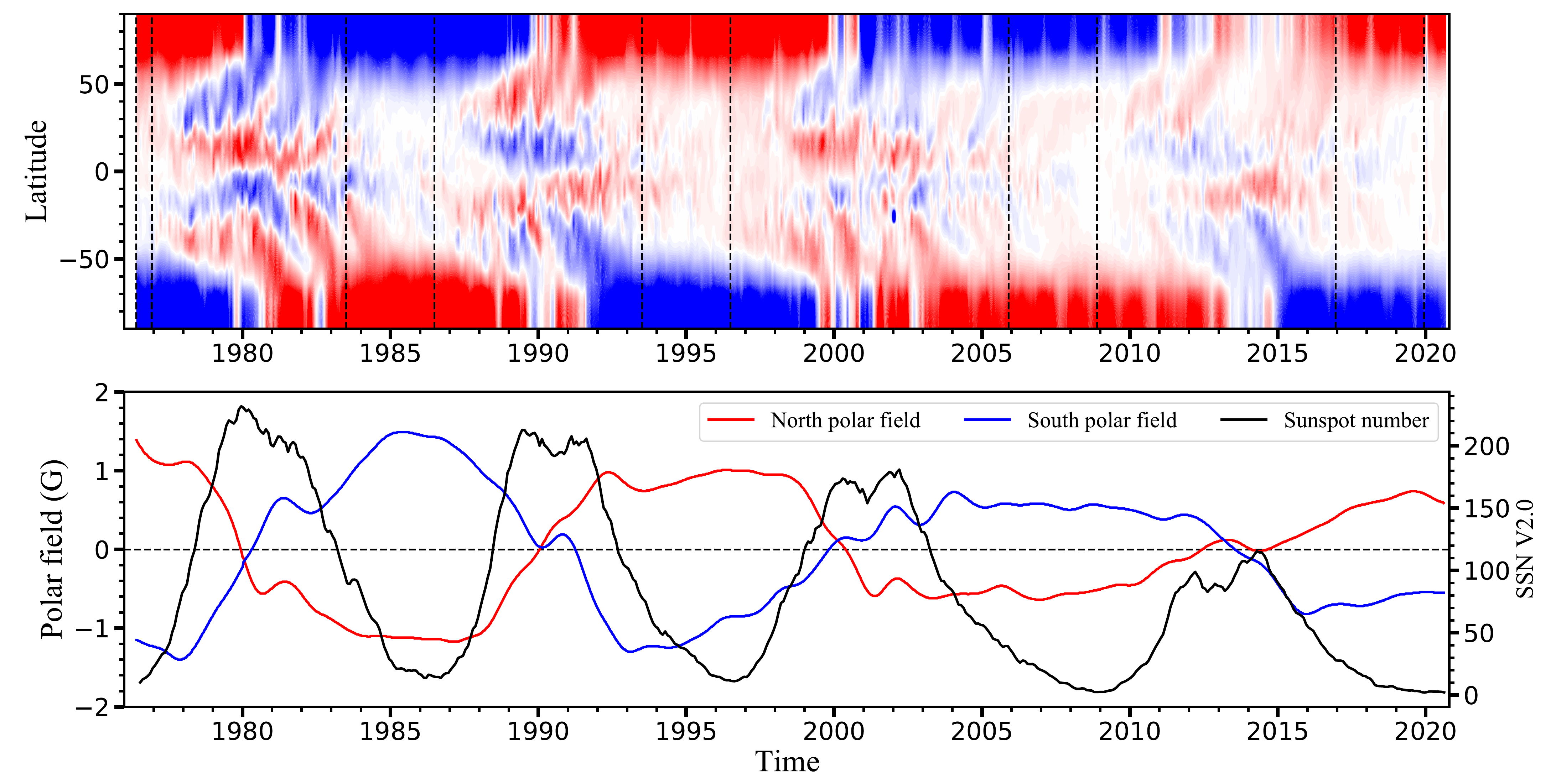}
\caption{Time evolution of the longitudinal averaged photospheric magnetic field (upper panel), sunspot number (lower panel, black curve), northern (red curve) and southern polar field (blue curve). The vertical dashed lines in the upper panel show the time span we choose as cycle's minimum.}
\label{fig:pf_sn_t}
\end{figure}

First, we multiply the observed data by a factor of 1.85 to correct the saturation effect. Another factor of 1.25 is particularly multiplied by the first two years data to correct the scattering effect. Then we divide the data by $\cos\theta$ to obtain the radial field at the surface with 30 data points in equal steps of sine latitude from +14.5/15 to -14.5/15. Next we use an extrapolation method to obtain the data of the polar field obeying the relation $\cos^{8} \theta$ \citep{Svalgaard1978}. After that,  we transform the data in equal steps of sine latitude to the values with equal latitude steps by the interpolation. Finally, 127 grids of data representing the radial field of the surface in equal latitude are obtained in accordance with the spatial resolution of the dynamo model. The corresponding time evolution of the longitude-averaged photospheric magnetic field $\langle B_r\rangle$ is shown in the upper panel of Figure \ref{fig:pf_sn_t}. The lower panel shows the time evolution of the sunspot number and the polar field. The polar field reverses around solar cycle maximum and peaks around cycle minimum. The polar field in Figure \ref{fig:pf_sn_t} is from WSO polar field measurements obtained from their website \footnote{http://wso.stanford.edu/Polar.html}, rather than calculated based on the WSO full-disc data presented in the upper panel.

Solar Cycle 25 Prediction Panel suggests that solar minimum may have occurred in December 2019 \footnote{https://www.weather.gov/news/201509-solar-cycle}, marking the start of a new solar cycle. See also \cite{Nandy2020}. Hence we take the timing of Cycle 24 minimum as 2020. The timings of the cycles 20-23 minima are based on the NGDC website \footnote{https://www.ngdc.noaa.gov/stp/space-weather/solar-data/solar-indices/sunspot-numbers/cycle-data/}. We use the averaged data among three years before each minimum except for Cycle 20 because of data shortage to make the prediction. To be more specific, data during CRs1642-1649 (1976:05:27-1976:12:04), CRs1737-1777 (1983:07:01-1986:06:26), CRs1871-1911 (1993:07:03-1996:06:28), CRs2037-2077 (2005:11:25-2008:11:20), and CRs2185-2225 (2016:12:14-2019:12:10) corresponding to the minimum periods of Cycles 20-24 are used. Besides, we also try the averaged data of one-year and two-year intervals before each cycle minimum. Around the time of polar field's maximum, we likewise perform a data assimilation to make the prediction. \cite{Svalgaard2005} and \cite{Svalgaard2020} used the averaged data of three-year intervals before each cycle minimum for their prediction model as well.

\begin{figure}[htbp]
\centering
\includegraphics[scale=0.32]{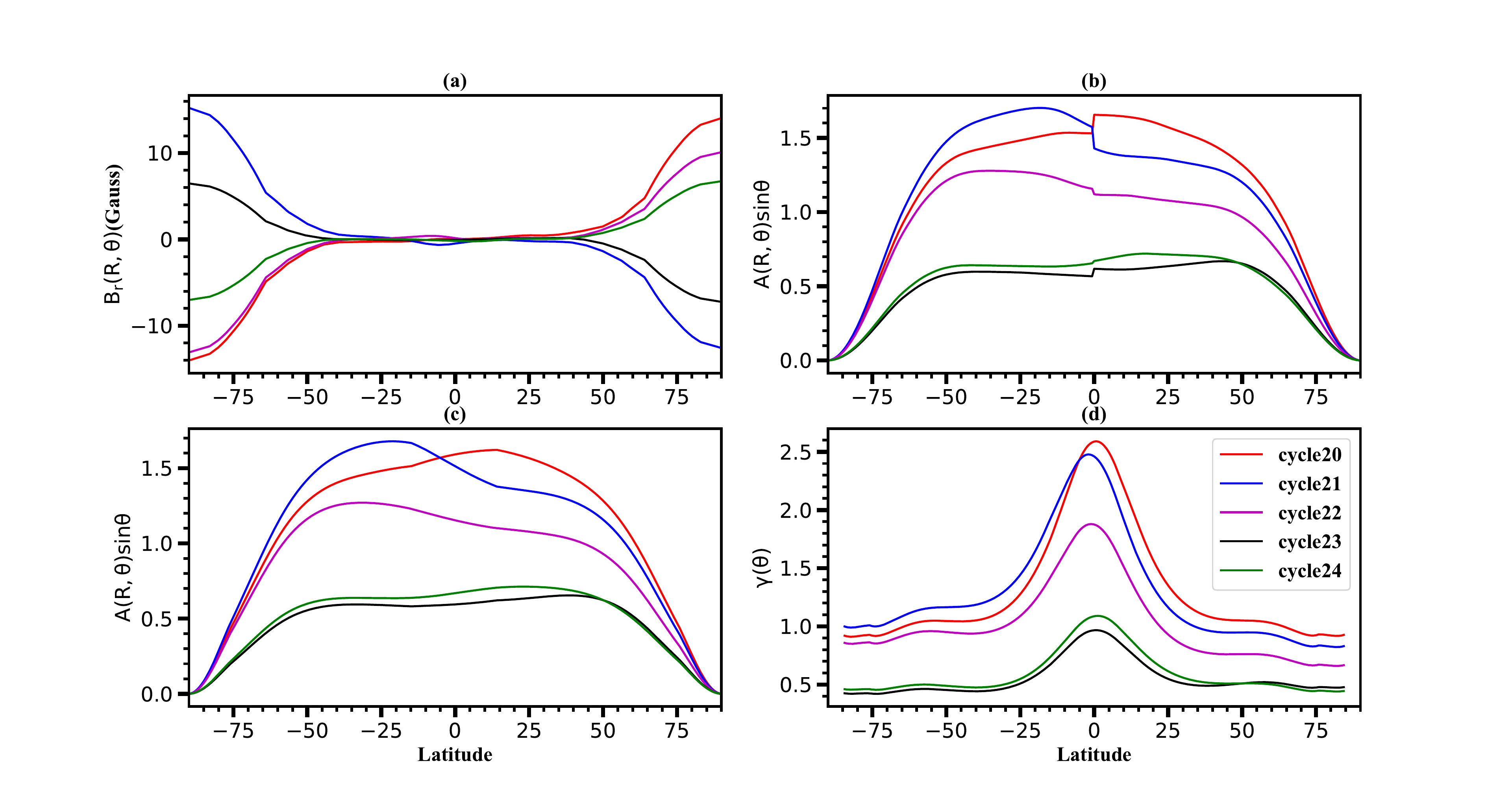}
\caption{Latitudinal dependence of the surface radial field $\langle B_r\rangle$ (a), unsmoothed (b) and smoothed (c) poloidal field $A\sin\theta$, and $\gamma({\theta})$ value (d) at each cycle's minima. The red, blue, purple, black, and green lines represent profiles of cycles 20, 21, 22, 23, and 24, respectively, in each panel.}
\label{fig:gama_t}
\end{figure}

In the following we present how to get the surface poloidal field $A\sin\theta$ from the observed surface $\langle B_r\rangle$ at each cycle minimum. Figure \ref{fig:gama_t}a shows the latitudinal distribution of $\langle B_r\rangle$ at the 5 cycle minima. We then apply the following integrations in the two hemispheres to obtain $A\sin\theta$
\begin{equation}A\left(\mathrm{R}_{\odot}, \theta, t\right) \sin \theta=\left\{\begin{array}{ll}
\int_{0}^{\theta} \langle B_r\rangle \left(\mathrm{R}_{\odot}, \theta^{\prime}, t\right) \sin \theta^{\prime} \mathrm{d} \theta^{\prime} & 0<\theta<\frac{\pi}{2}, \\
\int_{\pi}^{\theta} \langle B_r\rangle \left(\mathrm{R}_{\odot}, \theta^{\prime}, t\right) \sin \theta^{\prime} \mathrm{d} \theta^{\prime} & \frac{\pi}{2}<\theta<\pi.
\end{array}\right.\end{equation}
Figure \ref{fig:gama_t}b shows the plots of $A\sin\theta$ as a function of latitude at the five minima. Since the observed magnetograms cannot assure the divergence of the magnetic field over the whole surface, we see the jumps of $A\sin\theta$ values around the Equator. The smoothed results are shown in Figure \ref{fig:gama_t}(c). Cycle 23 provides the average cycle amplitude. Hence we take the poloidal field at its beginning, i.e., the averaged poloidal field of three-year intervals before the cycle 22 minimum, as the poloidal field of average strength corresponding to the result in that phase of the dynamo.

The latitudinal averaged value of $A\sin\theta$, $\overline{A\sin\theta}$, at Cycle 22 minimum is 0.871, while $\overline{A\sin\theta}$ from our standard dynamo model is 0.0047. Hence we use a proportional coefficient $c=$185.14 to calibrate between the observation and the simulation. In \cite{Jiang2007} the calibration is resorted to the critical magnetic field $B_c$. In essence the two ways are equal. Then a latitude-dependent factor $\gamma(\theta)$ is derived by the division of the observed $A\sin\theta$ over the value from the calibrated standard dynamo model for the five cycles minima. The values of $\gamma(\theta)$ for the 5 cycles are shown in Figure \ref{fig:gama_t}d. They will be used as the observed data fed into the dynamo model. Above 85$^\circ$ latitude the $\gamma(\theta)$ values are fixed to be 1 since both the observed and simulated $A\sin\theta$ are small as shown by \cite{Jiang2007}.

We assume that the poloidal field correction on the dynamo due to the BL mechanism is within the near surface layer from $0.85R_\odot$ to $R_\odot$. During every cycle minimum, we correct the dynamo simulation by multiplying $A\sin\theta$ the $\gamma(\theta)$ factor for every grid point above $0.85R_\odot$ and resume the simulation. This produces our data-driven prediction for Cycle 25.

\section{Prediction Results}\label{s:results}
\begin{figure}[htbp]
\centering
\includegraphics[scale=0.3]{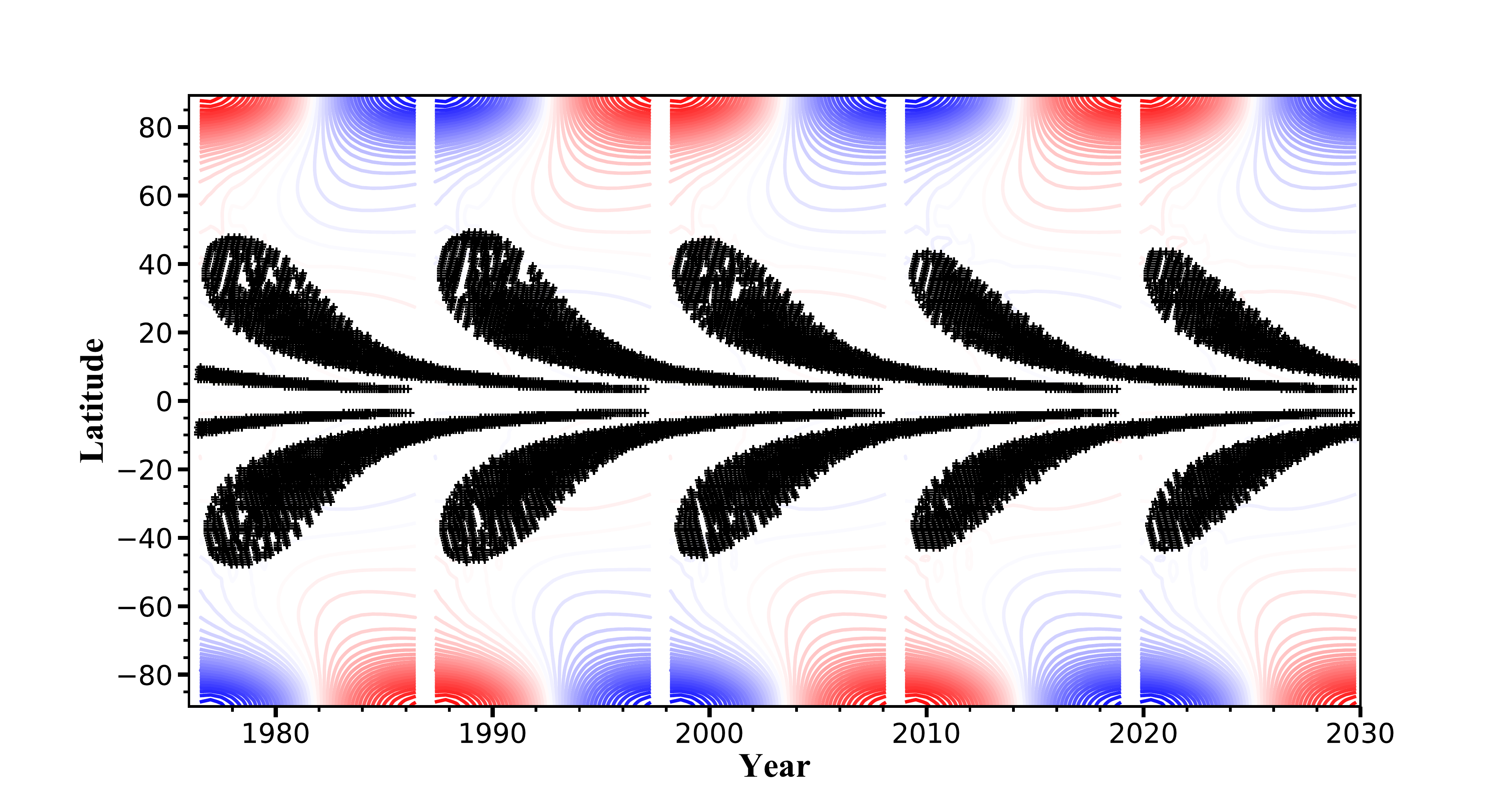}
\caption{Time-latitude diagram for the radial field at the surface and the distribution of the sunspots (denoted by symbol ``+'') from the dynamo-based prediction model. Contours of the radial field with red and blue colors represent the opposite polarities. The gaps correspond to the timings of data assimilations into the dynamo model.}
\label{fig:torPor_t}
\end{figure}

Figure \ref{fig:torPor_t} shows the time evolution of the sunspot emergence denoted by ``+" and the contour of radial field for Cycles 21-25 based on the data assimilation of flux transport dynamo modeling. The gaps correspond to the timings of data assimilations. The radial field has two branches. One shows an equatorward propagation at low latitudes. The other has a poleward propagation at high latitudes. The poleward branch shows the anti-phase with the sunspot emergence, which is confined within $\pm40^\circ$. Weaker cycles have lower latitudes emergence. All these properties are in conformity with the observations.

\begin{figure}[htbp]
\centering
\includegraphics[scale=0.28]{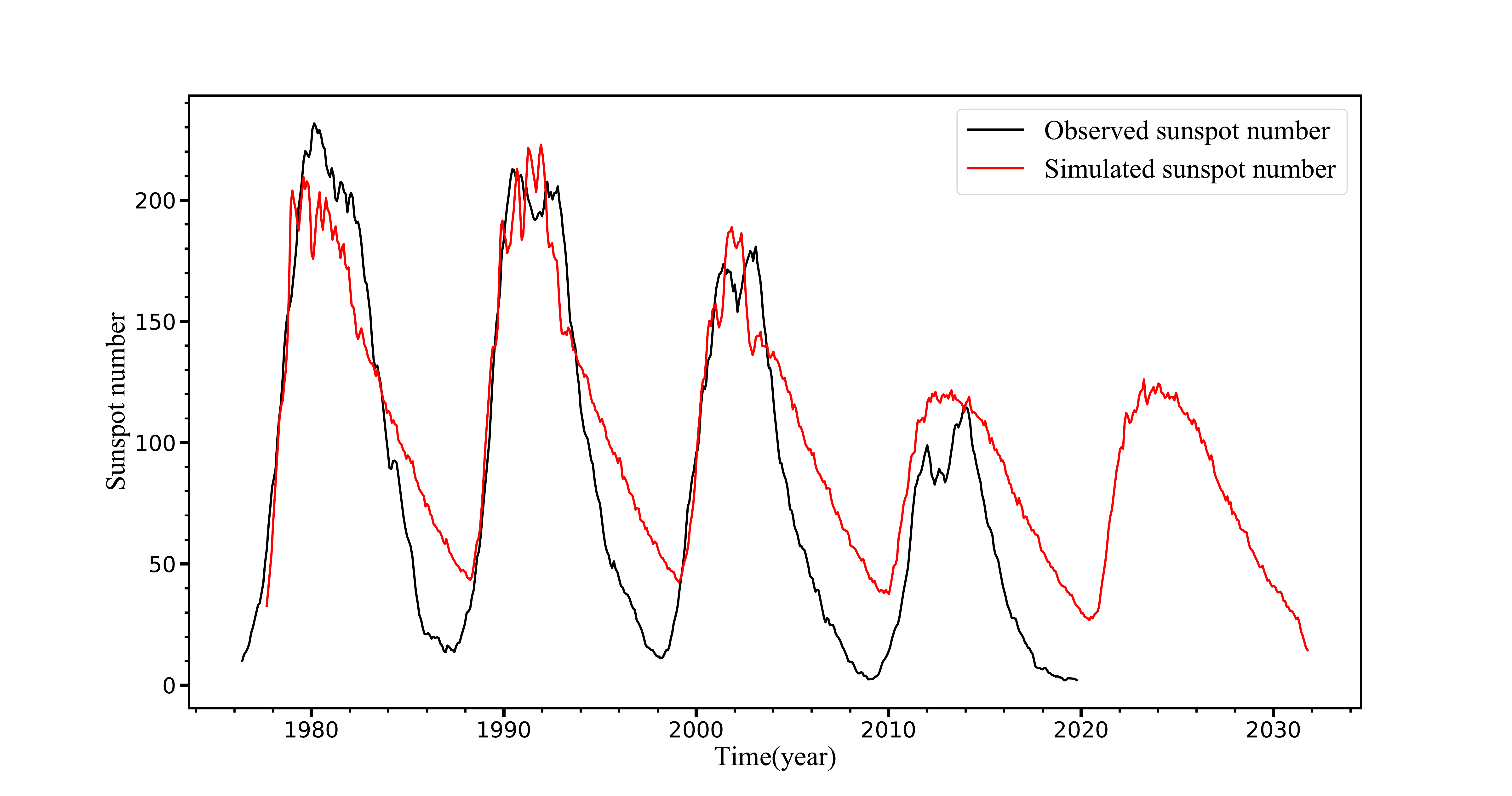}
\caption{Observed and simulated sunspot number as a function of time described by the black and red solid line, respectively. The observed solar cycle length is compressed or expanded to be consistent with the solar cycle length of the dynamo model, i.e., 10.84 yr.}
\label{fig:pre_sn_t}
\end{figure}

As mentioned above, the cycle period of our dynamo model has a fixed value of 10.84 years. The duration of each observed solar cycle is different. Especially cycle 23 has an extended minimum. The variable cycle periods from observations and the fixed cycle periods from the dynamo model may cause phase deviations between predictions and observations in the time series of solar cycle evolution. Therefore, we compress or expand the observed solar cycle length to be consistent with the fixed solar cycle length of the dynamo model. Figure \ref{fig:pre_sn_t} presents the time evolution of the monthly sunspot number generated from the prediction model (red curve) superposed on the observed one (black curve) after the adjustment of the cycle length. The predicted amplitudes of solar cycles 21-24 show a good agreement with the observed ones. The Pearson correlation coefficient between them is 0.96 with a confidence level of 95\%. Furthermore, the rising phases of cycles 21-23 are quite consistent between the observed and simulated data. The rising phase of cycle 24 has a slight shift between prediction and observation. This results from the extended cycle 23 minimum. The rising phases of the solar cycle correspond to the Waldmeier effect, which indicates that a stronger (weaker) cycle tends to have a shorter (longer) rising phase. The Waldmeier effect is a natural result of the BL-type dynamo with the magnetic memory less than one cycle \citep{Karak2011}. Under the effect of the strong diffusivity we adopt for the poloidal field, the surface poloidal field is transported into the tachocline in about 5 years. Then it takes the stronger (weaker) poloidal field less (more) time to be stretched and amplified to  reach the critical torodial field strength $B_c$ under the effect of the differential rotation, and hence to generate the quicker (slower) rise of solar cycles. The decline phase from the prediction model shows a large deviation from the observations. As suggested by \cite{Cameron2016}, the decline phase of the cycle is dominated by a diffusion process. The deviation might be an indication that some physics, e.g., the diffusivity of the toroidal field, in the dynamo model is not realistically modelled.

\begin{table}[t]
  \caption{Predicted cycle 25 amplitude by assimilating the longitudinal averaged WSO photospheric magnetic field at different timings. The correlation coefficient corresponds to the correlation between the predicted amplitudes of solar cycles 21-24 and the observed ones.}
\begin{tabular}{ccc}
\hline
     Timing of data assimilation  & Cycle 25 amplitude &  Correlation coefficient\\
\hline
    Average over 3 years before cycle mini. & 126.1 & $r$=0.96 ($p$=0.055)\\
    Average over 2 years before cycle mini. & 125.3 & $r$=0.97 ($p$=0.052)\\
    Average over 1 years before cycle mini. &  124.3 & $r$=0.96 ($p$=0.055)\\
    Timing of the maximum polar field & 120.2 & $r$=0.96 ($p$=0.055)\\
\hline
  \end{tabular}%
  \label{tab:tabl}%
\end{table}%


Figure \ref{fig:pre_sn_t} shows that the amplitude of upcoming Cycle 25 is 126.13, which is 9.4\% stronger than that of cycle 24. The cycle will reach its maximum around 2023 - 2024. The predicted amplitude of Cycle 25 is consistent with \cite{Svalgaard2020}, who took the magnitude of the polar field near the solar minimum as a precursor to make the prediction. The attempts of choosing the averaged surface poloidal field over one-year and two-year intervals before the cycle minimum as inputs of poloidal fields reach almost the same outcomes. The predicted Cycle 25 amplitudes are 124.3 and 125.3, respectively. We have made another experiment mentioned in Section \ref{sec:DataAssimilation} by assimilating the poloidal field at the timing of the maximum polar field. The Pearson correlation of the cycle amplitudes between the observation and the prediction is slightly higher. The prediction amplitude of Cycle 25 is 120.17. The results based on different data inputs are summarized in Table \ref{tab:tabl}, which shows that the predictions weakly depend on the timing of data assimilation. The reason for the weak dependence is due to the minor variation amplitudes of the polar field three years before each cycle minimum during the past four cycles, as shown in the low panel of Figure \ref{fig:pf_sn_t} and as presented by \cite{Iijima2017}. Please note that we cannot get the conclusion that sunspot emergence three years before each cycle minimum has trivial effects on the polar field at the cycle minimum. The stochastic properties of sunspot emergence could cause large effects on the polar field, especially for the weak cycles. This is clearly demonstrated by Figures 3g-3i of \cite{Jiang2020}, who investigate the axial dipole moment evolution by synthesizing the sunspot emergence of a large amount of solar cycles. The best timing of data assimilation for solar cycle prediction remains to be investigated. All predictions show that Cycle 25 would be slightly stronger than Cycle 24. Hence based on the dynamo-based model, the Sun may not be approaching a Maunder-like grand minimum. It could be a kind of global minimum similar to the phase of Solar Cycles 12-14.

\section{Conclusions and Discussion}\label{s:Conclusion}
In this article we have applied the first successful dynamo-based solar cycle prediction model developed by \cite{Jiang2007} to the prediction of Cycle 25. The result shows that the amplitude of Cycle 25 would be 126, which is 9.4\% stronger than that of Cycle 24. And the prediction does not support the arriving of a Maunder-type grand solar minimum. The development of the sunspot number during the next few years will verify whether the model included the correct physics of the solar cycle.

The physics responsible for the prediction skill consists of two aspects. One is that the surface poloidal field seeds for the subsequent cycle. The other is the cycle memory, which is about 5 years, so that the polar field around the cycle minimum determines the subsequent cycle strength. If the dynamo-based prediction is proven to be correct again, the prediction model will further support that the surface poloidal field is the source of the subsequent cycle. In return, this provides two constraints on the solar dynamo. One is the source of the poloidal field. It corresponds to the surface poloidal field due to flux emergence and evolution.The other is about the poloidal flux transport time from the surface to the toroidal field generation layer. It is less than one cycle. Still if the dynamo-based prediction is proven to be correct, it does not mean that the dynamo model we use is realistic. There are still some outstanding questions about the solar dynamo, for example, the roles of the tachocline and meridional circulation, the property of the turbulent diffusivity and so on. Helioseismology, 3 dimensional MHD simulations, solar-stellar connection, etc. could provide insight into these open questions. Updated 2D and 3D kinematic dynamo models remain to be developed. At present there are some attempts, e.g., \cite{Jiang2007b,Yeates2013, Miesch2014,Kumar2019}, to develop  3D kinematic dynamo models.

Our physics-based prediction has a fixed cycle period. Hence the model could just give a prediction of the cycle amplitude, but not the cycle shape. And it cannot provide physical insight into the variation of the cycle periods. \cite{Labonville2019} made the first attempt to apply the fully coupled surface poloidal field evolution with a flux transport dynamo to the solar cycle prediction. The model is supposed to have the ability to predict not only the cycle amplitude but also the cycle shape.
Other physics-based predictions, especially the application of SFT models to make earlier cycle predictions even before cycle minimum \citep[e.g.,][]{Cameron2016,Hathaway2016,Bhowmik2018,Jiang2018}, have also been emerging in recent years. These physics-based prediction models compared with the behaviour of Cycle 25 will help us understand the solar cycle better.

%
\begin{acks}
We thank Zi-Fan Wang, Ze-bin Zhang, and Bidaya Karak for reading this manuscript. Surya code is courtesy of Arnab Rai Choudhuri and his students who developed the code. We acknowledge the use of magnetic data from the Wilcox Solar Observatory. This research was supported by the National Natural Science Foundation of China through grant Nos. 11873023 and 11522325, the Fundamental Research Funds for the Central Universities of China, Key Research Program of Frontier Sciences of CAS through grant No. ZDBS-LY-SLH013, the B-type Strategic Priority Program of CAS through grant No. XDB41000000, and the International Space Science Institute Teams 474 and 475.
\end{acks}
\\
\\
\textbf{Disclosure of Potential Conflicts of Interest} The authors declare that there are no conflicts of interest.

\bibliographystyle{spr-mp-sola}
\bibliography{Reference}

\end{article}
\end{document}